\documentstyle[12pt]{article}

\newcommand{\de}{\partial}
\newcommand{\be}{\begin{equation}}
\newcommand{\ba}{\begin{eqnarray}}
\newcommand{\ea}{\end{eqnarray}}
\newcommand{\ee}{\end{equation}}

\newcommand{\beq}{\begin{equation}}
\newcommand{\eeq}{\end{equation}}
\newcommand{\beqa}{\begin{eqnarray}}
\newcommand{\eeqa}{\end{eqnarray}}

\newcommand{\th}{\theta}

\newcommand{\cD}{{\cal D}}

\newcommand{\bra}[1]{\left\langle\, #1\,\right|}
\newcommand{\ket}[1]{\left|\, #1\,\right\rangle}

\newcommand{\ra}{\rightarrow}

\newcommand{\bR}{{\bf R}}
\newcommand{\bZ}{{\bf Z}}
\newcommand{\Dslash}{D\!\!\!\!\slash\,}

\newcommand{\hX}{{\hat X}}

\newcommand{\ddbp}{\mbox{D}p-\overline{\mbox{D}p}}
\newcommand{\dz}{\mbox{D}0{\rm -}\overline{\mbox{D}0}}

\def\Tr{\mathop{\rm Tr}\nolimits}

\def\mat#1{\matt[#1]}
\def\matt[#1,#2,#3,#4]{\left(%
\begin{array}{cc} #1 & #2 \\ #3 & #4 \end{array} \right)}

\begin{document}
\begin{titlepage}
\thispagestyle{empty}
\begin{flushright}
hep-th/0701179 \\
YITP-07-01 \\
January, 2007 
\end{flushright}

\bigskip

\begin{center}
\noindent{\Large 
\textbf{Supertubes in Matrix model and DBI action}}\\
\vspace{2cm}
\noindent{
Seiji Terashima\footnote{E-mail: terasima@yukawa.kyoto-u.ac.jp}  
}\\

\vspace{.2cm}

%${}^1$
 {\it  Yukawa Institute for Theoretical Physics, Kyoto University\\ 
Kyoto 606-8502, Japan}
\vskip 2em
\bigskip

\end{center}
\begin{abstract}

We show the equivalence between the supertube solutions with
an arbitrary cross section
in two different actions, 
the DBI action for the D2-brane 
and the matrix model action for the D0-branes. 
More precisely, 
the equivalence between the supertubes in the 
D2-brane picture and the D0-brane picture
is shown in the boundary state formalism
which is valid for all order in $\alpha'$.
This is an application of 
the method using the infinitely many D0-branes and anti-D0-branes
which has been used to show 
other equivalence relations between two seemingly 
different D-brane systems, 
including the D-brane 
realization of the ADHM construction of instanton.
We also apply this method to 
the superfunnel type solutions successfully.

\end{abstract}
\end{titlepage}

\newpage

\tableofcontents

\section{Introduction and Summary}

There are (marginal) bound states of the D-branes with different dimensions
in string theory, which have been very important for studying string theory.
Furthermore, it have been known that 
such bound states have two (or several) different, 
but equivalent descriptions. The typical example 
is the bound state of the D4-branes and the D0-branes 
\cite{Wi, Do}.
For a bound state of D$p$-branes and D$q$-branes,
we can think the D$p$-branes as solitons in the D$q$-branes.
Instead, we can think the D$q$-branes as solitons in D$p$-branes.
This equivalence or duality between different D-brane systems is
one of the characteristic properties 
of the D-branes and will be important to investigate further.
However, it is in general very difficult to 
prove the duality 
although the evidences for the duality 
have been given for many bound states of D-branes
by using the low energy effective actions and supersymmetry.
The two dual D-brane systems
are very different in their low energy effective actions, for example,
the DBI action and the matrix model action.
Therefore, it will be impossible to show the duality by using the 
low energy effective actions.
Moreover, these two actions are only valid for
different regions in the parameter space
except in some large $N$ limit where $N$ is the number of D-branes.
Then we should heavily rely on the BPS protected properties
to connect two different regions of the parameter space. 
Using the boundary state formalism the duality can be shown, however,
only for some special cases \cite{Is, HiNoSu}.

Recently, a unified picture of such dualities
was given by the tachyon condensation of the unstable D-branes 
in the boundary state formalism \cite{Te2} \footnote{
This can be regarded as a generalization of \cite{Is}.}
and successfully applied to the bound state of 
the D4-brane and the D0-branes, i.e. ADHM(N) construction of 
instantons and monopoles \cite{HaTe3, HaTe4}. 
In this paper, we will consider this duality for the supertubes
by the method using the tachyon condensation \cite{Te2}.

The supertubes are the 1/4 supersymmetric bound states of the D0-branes 
with fundamental strings
which are expanded to the 
tubular D2-branes \cite{MaTo}. %\footnote{Here we consider
%the type IIA superstring theory,
%however, generalization to type IIB super string theory or
%to higher dimensional D-branes are straightforward.}
% which do not have D2-brane charge.
%we give the angular momentum to 
Therefore the supertubes have two different pictures:
the D0-branes and the D2-brane.
In the D0-brane picture the supertube solutions was found 
in the matrix model by \cite{BaLe}
and the supertubes have been 
investigated extensively in both pictures \cite{EmMaTo}-\cite{BaOhTo}.

In this paper we construct the supertube solution of the D2-brane
from the infinitely many D0-branes and anti-D0-branes.
Then by using the different basis of the Chan-Paton factor
of the $\dz$-branes, we find the supertubes in the D0-brane picture.
This implies that 
these two supertubes solutions in
the D2-brane picture and the D0-brane picture
are indeed equivalent in the boundary state formalism 
which is valid for all order in $\alpha'$.\footnote{
The boundary state of the supertube in the T-dual of D0-brane picture
was found in \cite{Ta} although we will not use it.}

%Note that 
%the D2-brane effective action is the DBI action,
%on the other hand,
%in the D0-branes picture the effective action is the matrix model.
%We also apply this method to 
%the superfunnel type solutions successfully.

Some remarks are as follows.
In this paper, we will consider the flat spacetime 
and
the classical limit $g_s \rightarrow 0$ only, thus 
we will drop the $g_s$ dependence for notational simplicity.
We will show the equivalence between 
the boundary states which can be off-shell.
The on-shell conditions can be derived in some approximations,
including the DBI actions and the matrix model actions.
%One important point here is these two are valid for
%different regions in the parameter space.
%except for some large $N$ limit where $N$ is the number of D-branes.

%We will show the two seemingly different 
%systems in the different descriptions 
%are indeed same classically.
%This means that 
%these two systems should be equivalent quantum mechanically
%since just the descriptions are different.
%Actually, we will show the two boundary sates are equivalent,
%which implies the loop corrections are equivalent.

Organisation of this paper is as follows. In section \ref{d0d0}
we review how to obtain the duality or equivalence 
relations between two different D-brane system 
from the D-brane-anti-D-brane system.
Section \ref{elec} is for the brief review of the 
supersymmetric solitons with electric field.
Then the method explained in section \ref{d0d0} is
applied to the supertubes with compact cross section
in section \ref{sd} and 
with non-compact cross section 
in section \ref{gene}.
The superfunnel type solutions are also discussed
in section \ref{gene}.
In the appendix we show the duality for the non compact 
supertube from the 
D2-brane-anti-D2-brane system.

%\section{Supertube Solution of the D2-brane and the D0-branes}

\section{Duality between D0-branes and Dp-brane via Tachyon}
\label{d0d0}

In this section we will review how to obtain 
the duality between Dq-branes and Dp-brane 
using the Dp-branes and the anti Dp-branes \cite{Te2},
in particular, 
the duality between D0-branes and Dp-brane in type IIA string theory
using the $\dz$-brane pairs, which will be applied to
the supertubes.

\subsection{Diagonalization of the tachyon and the duality}

Consider $N$ D$p$-branes and $N'$ anti-D$p$-branes.
They have the tachyon $T'$ which is a complex $N \times N'$ matrix.
Note that 
the Chan-Paton bundle
of the $N$ D$p$-branes is an $N$ dimensional vector space
on which $T'$ acts.
For the $N'$ anti-D$p$-branes, 
there is an $N'$ dimensional vector space
on which $T'^\dagger$ acts.
Any orthonormal basis of the two vector spaces can be used and  
the basis change is the $U(N) \times U(N')$ gauge symmetry on
the D-branes and the anti-D-branes.
Then, generally, 
by giving topologically nontrivial expectation value to the tachyon,
we get the D$q$-branes with $q \neq p$, 
i.e. the decent relation \cite{Sen, Senreview} for $q<p$
or the accent relations for $q>p$ \cite{Te2, AsSuTe1}.
Thus after the tachyon condensation 
we can identify the system of the D$p$-branes and 
the anti-D$p$-branes 
as
the D$q$-branes (with surviving D$p$-branes or anti-D$p$-branes
after the tachyon condensation).

%On the other hand, we can diagonalize $T'$ 
%by the gauge transformation.
Now let us choose another basis of the Chan-Paton bundle
in which the tachyon $T'$ is diagonal.
In this gauge, we expect that 
the D$p$-branes (and the anti-D$p$-branes) corresponding 
to the nonzero eigen value of the $T'$ (and $T'^\dagger$) 
disappear, respectively. 
Then,
only the D$p$-branes (or the anti-D$p$-branes) 
corresponding to the zero modes of $T'$ 
(or $T'^\dagger$) will remain after the tachyon condensation,
respectively.
Actually, in the boundary state formalism 
or the boundary string field theory,
%the eigen value of the $T'$ or $T'^\dagger$ 
%will be infinite 
this is justified \cite{KMM2, KrLa, TaTeUe}.
Therefore the system of the D$p$-branes and 
the anti-D$p$-branes with the nontrivial tachyon condensation
is equivalent to 
D$p$-branes or anti-D$p$-branes corresponding to the zero modes
of the tachyon.
Note that we had 
the two different results of the same tachyon condensation
with the different gauge choices.

Thus, combining these two equivalences, 
we have an equivalence or a duality between the 
the D$q$-branes and 
D$p$-branes (or anti-D$p$-branes corresponding to the zero modes)
\cite{Te2}.
This was explicitly performed for
the flat noncommutative D-brane \cite{El, Te2}, 
fuzzy sphere \cite{Te2} and
ADHM(N) construction of the instantons and
monopoles \cite{HaTe3, HaTe4}.
In the next subsection 
we will explain this duality for $p=0$ case
in detail.

\subsection{Duality between infinitely many D0-branes and Dp-brane}
\label{d0d0a}

Consider $N$ $\dz$-brane pairs and let 
a D0-brane and an anti-D0-brane of each pair
are at a same position.
Then, the fields on the pairs are 
$N \times N$ matrix coordinates $X^\mu$ and
the complex tachyon $T'$, 
other than massive fields which are not relevant
in this paper.
Instead of $T'$, 
we will use $2N \times 2N$ Hermitian matrix
$T=\mat{0,T',T'^\dagger,0}$ for convenience.
%Note that 
%the $N$ D-branes span an $N$ dimensional vector space
%of the Chan-Paton index on which $T'$ acts.
%The $N$ anti-D-branes also span an $N$ dimensional vector space
%on which $T^\dagger'$ acts.
%Any orthonormal basis of the two vector space can be used and  
%the base change is the $U(N) \times U(N)$ gauge symmetry on the 
%$N$ $\dz$-brane pairs.

Now, let us consider a large $N$ limit, then
the matrices $X^\mu$ and $T$ become
operators acting on a Hilbert space.
The important configuration of the $N$ $\dz$-brane pairs
is
\beqa
T=u \Gamma^i (\de_i+i A_i(x)), \;\;\; 
\hat X^i=x^i, \;\;\; (i=1,\cdots,p)
\label{sol1}
\eeqa
where the Hilbert space is spanned by the 
(normalizable) spinors on $\bR^p$,
$\Gamma^i$ is the gamma matrix on $\bR^p$,
$ A_i(x)$ is a arbitrary function of $x^j$
and $u$ is just a constant which is sent to infinity.
In the $u \ra \infty$ limit,
the configuration (\ref{sol1}) 
represents a flat D$p$-brane with 
the background gauge field $A_i(x^i)$ and 
it becomes the exact solution of the equations of motion
of the boundary string field theory for the constant field strength
\cite{Te, AsSuTe1, AsSuTe2}.
(If $u$ is finite, (\ref{sol1}) does not corresponds to the 
solution of the equations of motion and is expected to represent
the D$p$-brane with some off-shell excitations.) 
This is ``T-dual'' to 
the well known decent relation \cite{Senreview}, 
which state that
we can have a D$q$-brane 
by the nontrivial tachyon condensation on
$\ddbp$-brane pairs where $q<p$.\footnote{
Both the decent relation and the accent relation (\ref{sol1})
are obtained from $[{\cal D}_i, T]=u \Gamma^i$ and 
$[{\cal D}_i, {\cal D}_j]=0$ by some truncation.
However, this is not the T-dual each other, of course.
The accent relation is also regarded as a generalization 
of the matrix model construction of a D$(2p)$-brane
using the noncommutative coordinates.}
Moreover, we can give the expectation value to the 
remaining transverse scalars, $\hat X^a$, on the $\dz$-branes 
which should be considered as an operator acting on the spinors: 
$\hat X^a=\phi^a(x^i)$ $(a=p+1, \cdots, 9)$
where $\phi^a(x^i)$ is an arbitrary function of $x^i$.
Then the D$p$-brane has the transverse scalars 
$X^a=\phi^a(x^i)$.

Here it is important to note that
the tachyon is proportional to the Dirac operator
on the world volume of the D$p$-brane, 
\beq
T=u \; \Dslash,
\eeq
and the D0-branes and anti-D0-branes correspond 
to the spinors with positive and negative chirality respectively.

Now assuming the ``Hamiltonian'' ${\cal H} \equiv \Dslash^2$ 
has a gap above the ground states or zero modes
\beq
{\cal H} \ket{a} =0, \;\; (a=1, \cdots,n)
\eeq
where we define the number of the zero modes as $n$.
Then we can follow the argument in the previous subsection:
Because $T^2=u^2 \Dslash^2 \ra \infty $ for nonzero modes,
the tachyon condensation forces $\dz$-brane pairs corresponding 
to the nonzero modes disappear and 
only the D0-branes and the anti-D0-branes corresponding 
to the zero modes survive.
Then the $n \times n$ matrix coordinates for the zero modes,
$\hat X^\mu_0 $,
will become
\beq
\left(\hat  X^\mu_0 \right)_{ab}
=\bra{a} \hat X^\mu \ket{b}, \;\;\; (\mu=1, \cdots,9)
\eeq
which is in general non-commutative although 
the infinite dimensional matrix $\hat X^i(=x^i)$ and $\hat X^a$
are commutative each other.
Note that 
we have only D0-branes
or only anti-D0-branes depending on the sign of 
the background field strength on the D$p$-brane, for a generic $T$,
i.e. all the zero modes are either positive chirality 
or negative chirality.
We will consider these cases only below.
Thus we have the equivalence or duality between
the D$p$-brane with the back ground gauge field $A_i$
and the $n$ D0-branes (or anti-D0-branes) with
the coordinates $\hat X^\mu_0 $.

It is easy to consider the D$p$-brane with
a curved world volume which is specified 
by the embedding $X^\mu=X^\mu(x^i)$.
Taking the tachyon as a Dirac operator with
the vielbein consistent with the induced metric,
we have the D$p$-brane after the tachyon condensation
\cite{AsSuTe1, Te2}.
Then, repeating the procedures for the flat D$p$-brane,
i.e. computing the matrix elements of the $\hat X^\mu$
between the zero-modes of the Dirac operator,
we have the D0-brane picture which should be equivalent
to the D$p$-brane picture.

It is important to note that
all the results in this paper can be
given and justified in the boundary state formalism \cite{AsSuTe3,Te2}
where the fields, like the tachyon, are represented 
in the world sheet boundary action,
although we will not explicitly write
the corresponding boundary states.\footnote{
We will use the naive off-shell extension of 
the boundary state, i.e. just inserting the 
possibly non-conformal world sheet boundary action
to the (on-shell) boundary state.
In this paper, the boundary state means such a naively
extended boundary state.}
Thus, the equivalences given in this paper 
are exact in all order in $\alpha'$.\footnote{
In this paper we take $\alpha'=2$.} 
We also note that the boundary state includes
any information about the D-brane, 
thus the equivalences in this paper imply the 
equivalences between the tensions, 
the couplings to the closed strings
and the D-brane charges,
of the D$p$-brane and the D0-branes.\footnote{
In particular, the equivalence between 
the D0-brane charges are reduced to the index theorem
\cite{AsSuTe3}. Actually, 
the three representations of the index 
\beq
 \int_{D_p} \Tr e^{F} \, \hat{\cal A}, \;\;\;\;
\Tr_{\rm zero \, modes \, of \, \Dslash}(\Gamma), \;\;\;
\Tr\left(\Gamma \, e^{-u^2 \Dslash^2} \right), \;\;\;
\eeq
correspond to the D2-brane, the D0-branes
and the $\dz$-brane pictures, respectively.
}

%The parameter $1/u$ enter the boundary action
%as a regularization. 
%We can include a regularization parameter $1/U$,
%then it becomes $(u+U)/(uU)$.
%Then we can take $u \rightarrow \infty$ limit first,
%which means $(u+U)/(uU) \rightarrow 1/U$.
%Then take the $U \rightarrow \infty$ limit.

\section{Supersymmetric Solitons with electric field}
\label{elec}

Let us consider the BFSS matrix model \cite{BFSS} which is same as 
the ``low energy'' effective action of the $N$ D0-branes 
in type IIA theory.
The equation of motion 
\beq
-\cD_t^2 \hat X^\mu 
+[\hat X_\nu,[\hat X^\nu,\hat X^\mu]]=0 \;\;\; (\mu=1, \dots,9)
\eeq
and Gauss low constraint
\beq
[\hat X_\mu, \cD_t \hat X^\mu]=0
\eeq
are simplified for time independent configurations with  
\beq
\hat A_t=\pm \hat Z, \;\;\; (\hat Z \equiv \hat X^9)
\eeq
to 
\beq
[\hat X_i, [\hat X^i, \hat X^\mu]]=0 \;\;\; (i=1,\dots,8).
\label{eom1}
\eeq
From this form, we can easily find two types of solitons:
the supertube type and the superfunnel type.

Introducing the operators $\hat a$ and $\hat N$ acting on 
the space spanned by the ket $\ket{m}$ such that
\beq
\hat{a} \ket{m} =\ket{m-1}, \;\;
\hat{N} \ket{m} =m \ket{m}, 
\label{aN0}
\eeq 
where $m \in \bZ$,
the supertube solution \cite{BaOhSh} is given by
\beq
\hat Z=\hat{N}-\varphi, \;\;  \hat X^i=\hat X^i(\hat{a}, \hat{a}^\dagger ),
\eeq
where $X^i(a, a^\dagger)$ is an arbitrary functions of 
$a, a^\dagger$ and  $0 \leq \varphi <1$.
We can check this is the solution of (\ref{eom1})
by using 
the following relations derived from the definition (\ref{aN0}):
\beq
[\hat{N},\hat{a}]=-\hat{a}, 
\;\;\;\; \hat{a}^\dagger=\hat{a}^{-1}, 
\;\;\;\; [\hat{a}^\dagger,\hat{a}]=0, \;\;\;\;
[\hat{N},f(\hat{a})]=-\frac{\de f(\hat{a})}{\de \ln \hat{a}},
\label{aNr}
\eeq
where $f(a)$ is an arbitrary function of $a$.
Note that $ \hat a^\dagger=\hat a^{-1}$ implies
$\hat a^\dagger \hat a=1$. 
For the circular case, i.e. $\hat X^1+i \hat X^2=R \; \hat{a}$ 
and others$=0$, 
this is the solitons obtained in \cite{BaLe}.
For a general profile, we will show later that 
this indeed represents the supertube 
with a general cross section.\footnote{
For a non-compact world volume,
the supertube is given by 
$\hat Z=\hat{z}$ and $\hX^i=\hX^i(\hat x)$
where $[\hat x, \hat{z}]=i/B$
and $B$ is a constant.}

The other type of solution is
given by 
\beq
\hat Z=\hat Z(\hat X^b), \;\;\; [\hat X^b, \hat X^c]=i (1/B)^{bc},
\;\;\; \;\; (b,c=1,\dots,2p)
\eeq
with 
\beq
[\hX_b,[\hX^b,\hat Z(\hX^c)]]=0.
\label{lap}
\eeq
If we take $B_{2i-1, 2i}=-B_{2i, 2i-1}=b_i \; (i=1,\dots,p)$
and others$=0$,
the eq. (\ref{lap}) becomes 
\beq
\triangle \hat Z(\hX) \equiv 
\sum_{i=1}^p \frac{1}{b_i^2}
\left( \frac{\de^2 }{\de^2 \hX^{2i-1}} 
+\frac{\de^2 }{\de^2 \hX^{2i}} \right) \hat Z(\hX)=0,
\eeq
where $\hat Z(\hX)$ is Weyl ordered.
Here we have used the fact that 
the derivative of the Weyl ordered product is 
again Weyl ordered.
Using the $\triangle G(X^i)=\delta (X^i)$,
we can write the solution as
\beq
\hat Z(\hX^i)=Z_0+\sum_{a} Q_a G(\hX^i-x^i_a),
\label{solbi}
\eeq
where $Q_a$ will be quantized by the flux quantization condition.
For $p=1$, this solution corresponds to 
the superfunnel of \cite{BaOhTo}.
In particular for 
$\hat Z(\hX^1,\hX^2)=Q/2 \log ((\hX_1)^2+(\hX_2)^2)$,
this solution have been obtained in \cite{BaOhSh, Hy} explicitly.
For a general $p$, this corresponds
to the BIon \cite{CaMa, Gi} with magnetic field.
Actually, it is well-known that 
by expanding around $\hat Z=0$ and $[\hX^i,\hX^j]=i \Theta_{ij}$,
the action of the D0-branes becomes 
the noncommutative D$(2p)$-brane action 
with magnetic fields \cite{Seiberg}.
Thus the solution (\ref{solbi}) corresponds 
to the noncommutative BIon.

Note that 
we could consider a combination of the supertube type and the superfunnel 
type of the solutions.

We note that for the D$p$-branes instead of D0-branes
the equations of motion (\ref{eom1}) is valid if we think
$X_k$ as $\cD_k \equiv \de_k+A_k$ for $k=1,\dots,p$.
and take $\de_t+i A_0=i Z$.
%The solution includes the BIon
%and the dyonic supertubes.

Now, following \cite{BaOhSh}, we will discuss the 
BPS equations of the matrix model.
The supersymmetric variation of the fermions in the matrix model is
\beq
\delta \psi= \left( 
\cD_t \hX^\mu \gamma_\mu
+\frac{i}{2} [ \hX^\mu, \hX^\nu] \gamma_{\mu \nu}
\epsilon+\epsilon'
\right).
\eeq
For the time independent configurations with $\hat A_t =\hat Z$,
we have the BPS equations
\beq
\delta \psi= \left( 
i [ \hX^i, \hX^j] \gamma_{i j}
P_- \epsilon+\epsilon'
\right),
\eeq
where we have defined the real projection operator 
$P_\pm=(1 \pm \gamma_z)/2$ 
and set $P_+ \epsilon=0$.
For $[\hX^i,\hX^j] = i \Theta^{ij}$ where $\Theta^{ij}$
is a constant,
we can solve it by $\epsilon'=
-i [ \hX^i, \hX^j] \gamma_{i j}P_- \epsilon$.
Therefore, the time-independent solution of the Gauss law constraint
(\ref{lap}) with $\hat A_t=\hat Z$ and $[\hX^i,\hX^j]=i \Theta^{ij}$
is indeed the 1/4 BPS solution of the matrix model.

\section{Supertube and the Duality}
\label{sd}

\subsection{Circular supertube}

In this subsection
we consider a supertube of a circular cross section 
which is a simplest supertube of a compact cross section 
in the flat spacetime. %\footnote{
%We will consider the duality for the supertubes 
%with arbitrary cross sections in Appendix A.}
In particular, we consider 
a D2-brane in type IIA superstring theory
whose world volume coordinates are labelled by 
$(t,z,\theta)$ with 
a constant field strength $F=E dt \wedge dz +B dz \wedge d \theta$. 
We will take a gauge,
\beq
A_z=0, \;\; A_t=-E z, \;\; A_\theta=B z + \varphi,
\eeq
where $\varphi$ is 
a Wilson line which can be taken as $0 \leq \varphi < 1$.\footnote{
Note that the space-time is $R^{1,9}$ and the embedding of 
the $S^1$ associated to the Wilson line into the space-time 
is contractable. 
Thus, even if we compactify the space-time to a tori
in $X^i$ directions,
this Wilson line does not corresponds to 
the position in the T-dual picture.}
Although we have taken a particular gauge,
the Wilson line has a gauge invariant and physical meaning.
The configuration of the D2-brane in the flat spacetime is 
$X^0=t, X^1=R \cos \theta, X^2=R \sin \theta, X^3=z$ and $X^i=0$ 
for $i=4,\ldots, 9$
and the induced metric on the D2-brane
is given by $ds^2=-dt^2+dz^2+R^2 d\theta^2$. 
If we take $E=\pm 1$, this configuration becomes BPS and 
satisfies the classical equations of motion 
of the DBI action \cite{MaTo}.

Now we will represent this D2-brane configuration
as a soliton of infinitely many $\dz$-brane pairs 
as explained in the subsection \ref{d0d0a}.
The tachyon of the $\dz$-brane pairs is 
\beq
T=u \Dslash,
\eeq
where $\Dslash$ is the Dirac operator on a time slice of 
the world volume of corresponding D2-brane 
(i.e. on the $\bR \times S^1$),
\beq
i \Dslash=\sigma_1 \frac{1}{R} 
\left( \frac{\de}{\de \theta} -i (Bz+\varphi) \right) 
+\sigma_2 \frac{\de}{\de z}.
\label{Dirac}
\eeq
The transverse scalars of the D0-anti D0-brane pairs 
are given by
\beq
\hat X^1=R \cos \theta, \; \hX^2=R \sin \theta, 
\; \hX^3=z, \; \hX^i=0 \;\;(i=4, \ldots,9),
\eeq
and
\beq
\hat A_t=-E z,
\eeq
where the tachyon, the scalars 
and the gauge potential, which are $N \times N$ matrices for 
the $N$ pairs, become operators acting on the 
Hilbert space spanned by the normalizable spinor valued functions on 
the $\bR \times S^1$ parametrised by $\{ z,\theta \}$
by taking a particular large $N$ limit.
Note that the information of 
the gauge potential of the D2-brane except
$\hat A_t$ are encoded in the tachyon of the D0-anti D0-brane pairs.
Then, these $\dz$-brane pairs are equivalent to the supertube 
of the D2-brane
(in the $u \ra \infty$ limit) \cite{Te, AsSuTe1}.

In order to obtain 
the matrix model picture 
(or more precisely the D0-brane picture), 
we need
to find the zero modes of the Dirac operator
in the $\dz$-brane picture,
which corresponds to
the D0-branes remain after the tachyon condensation.
Thus what we should solve is 
\beq
\Dslash \Psi(\theta,z)=0.
\label{diraczero}
\eeq
Because $\theta$ is periodic, we expand $\Psi$ as
$\Psi(\theta,z)=\sum_{m=-\infty}^{\infty} e^{i m \theta} \tilde{\Psi}_m(z)$.
Then we have
\beq
\left(  1_{2 \times 2} \frac{1}{R} (m-B z -\varphi )
+\sigma_3 \frac{\de}{\de z}  \right) \tilde{\Psi}_m(z) =0,
\eeq
where we multiplied $-i \sigma_1$ 
to the zero-mode equation (\ref{diraczero}).
Then for $B>0$ we find the zero-modes $\Psi_m(\theta,z)$ parametrised 
by an integer $m$,
\beq
\Psi_m(\theta,z)=\left( \frac{R}{4 \pi B} \right)^\frac{1}{4} 
\exp \left( -\frac{B}{2R} \left( z-\frac{m-\varphi}{B} \right)^2 
-i m \theta \right)
\left(
\begin{array}{cc} 0 \\ 1  \end{array} \right)
\eeq
and
for $B<0$,
\beq
\Psi_m(\theta,z)=\left( \frac{R}{4 \pi |B|} \right)^\frac{1}{4} 
\exp \left( -\frac{|B|}{2R} \left( z-\frac{m-\varphi}{B} \right)^2 
-i m \theta \right) 
\left(
\begin{array}{cc} 1 \\ 0  \end{array} \right)
\eeq
where $\Psi_m$ is normalized such that,
\beq
\int d\theta dz \Psi_m^\dagger \Psi_n =\delta_{mn}.
\eeq
Note that the chirality operator $i\sigma_1 \sigma_2 =-\sigma_3$ distinguish
the D0-brane and the anti-D0-brane.
We will consider $B>0$ case below without loss of generality.
Then the matrix elements of 
$\hX^\pm \equiv \hX^1 \pm i \hX^2=R e^{\pm i \th},\hX^3=z$ 
and $\hat A_t=-E z$
between the zero-modes
are computed as
\beqa
(\hX^+_0)_{mn} &=& \int d\theta dz \Psi_m^\dagger(\th,z) 
R e^{i \th} \Psi_n (\th,z) 
=R e^{-\frac{1}{4 B R}} \delta_{n,m+1}, \;\;
(\hX^-_0)_{mn} =R e^{-\frac{1}{4 B R}} \delta_{n+1,m}, \nonumber \\
(\hX^3_0)_{mn} &=& \int d\theta dz \Psi_m^\dagger(\th,z) 
z \Psi_n (\th,z) 
=\frac{n-\varphi}{B} \delta_{m,n}, \nonumber \\
(\hat A_{t,0})_{mn} &=& -E \, (\hX^3_0)_{mn}.
\label{vev}
\eeqa

Therefore, in the $u \rightarrow \infty$ limit, 
we have the D0-branes parametrised by an integer $m$
with the matrix coordinates $(\hX^i_0)_{mn}$ and 
a gauge field $(\hat A_{t,0})_{mn}$.
As explained in the previous section, 
this system of the infinitely many D0-branes 
should be equivalent to the D2-brane.
Note that from (\ref{vev}) we can see that 
the $n$-th D0-brane is at $z=\frac{n-\varphi}{B}$,
however, it does not have a definite position for the $x$ and $y$
directions because of the noncommutativity except in the 
$B \rightarrow \infty$ limit.
The nonzero Wilson line in the D2-brane picture 
corresponds to the 
shift of the D0-branes along the $z$ axis in the D0-brane picture.
We also see that the density of D0-branes 
along $z$ is $B$ as expected. 
Note that $E$ has mass dimension 2, but
$B$ has mass dimension 1 from the definition 
$F=E dt \wedge dz +B dz \wedge d \theta$. 

Introducing the operators $\hat a$ and $\hat N$ acting on 
the space spanned by the ket $\ket{m}$
such that
\beq
\hat a \ket{m} =\ket{m-1}, \;\;
\hat N \ket{m} =m \ket{m}, 
\label{aN2}
\eeq 
where $\ket{m}$ corresponds to the remaining $m$-th D0-brane,
we can view the matrices as the operators acting on this space:
\beq
\hat{X}_0^+= R e^{-\frac{1}{4 B R}} \, \hat a^\dagger, \;\;
\hat{X}^-_0= R e^{-\frac{1}{4 B R}} \, \hat a, \;\;
\hat{X}^3_0= \frac{\hat N-\varphi}{B}, \;\;
\eeq
which satisfy
\beq
[\hat{X}^3_0, \hat{X}^\pm_0] = \pm \frac{1}{B} \hat{X}^\pm_0,
\;\;\;\; [\hat{X}^+_0, \hat{X}^-_0] =0.
\label{com}
\eeq
From the relation following from (\ref{aNr}),
$\hat a^\dagger \hat a=1$,
we find
\beq
\hat{X}^+_0 \hat{X}^-_0= (\hat{X}^1_0)^2+ (\hat{X}^2_0)^2 
=R^2 e^{-\frac{1}{2 B R}},
\label{casimir}
\eeq 
is the Casimir.

We note that 
although the zero-mode $\Psi_m(\theta,z)$ was identified as $\ket{m}$,
in the reduced Hilbert space
we can also represent $\ket{m}$ as $e^{i \theta m}$
and $\{ \hat a, \; \hat N \}$ 
as $\{ e^{- i \theta}, \; \frac{1}{i} \frac{\de}{\de \theta}-\varphi \}$.

It is very interesting to note that
(\ref{com}) solves the equations of motion of 
the matrix model for the BPS D0-branes.
Actually, this is same as the solution of the matrix model
equations of motion given in \cite{BaLe}
although the parametrisation of $R$ is different. 
%Because the matrix model action is valid 
%when the DBI action of D2-brane is not in general valid,  
Therefore we have shown the equivalence between
the supertube of the D2-brane picture and the supertube 
of the D0-brane picture.

In the D2-brane picture 
we took the solution of the BPS equations of the DBI action.
Because the DBI action neglects the higher derivative terms
and the matrix model action neglects the higher order terms,
the BPS equations of the DBI action 
should be different from the BPS equations of the matrix model.
Thus we expect the supertube solution of the DBI action
is BPS even if 
we include $\alpha'$ corrections
since the both BPS equations are satisfied at the same time.

We have seen that the D0-branes
with (\ref{vev}) is essentially same as 
the solution of the matrix model
of \cite{BaLe} which was supposed to be the same as the D2-brane.
Here we stress that 
the correspondence between the 
D2-brane and the D0-branes are exact in our approach.
One of the important points of the correspondence is that 
the naively defined radius of the supertube are not same for the
two equivalent systems.
Actually, the square root of (\ref{casimir}) could be considered as 
the radius of the supertube, 
but it is smaller than $R$ 
by a factor of $\exp (-\frac{1}{2 BR} )$.
We will discuss the meanings of this difference 
in the subsection \ref{D0D2}.

\subsection{General cross section}
\label{general cross section}

Let us consider general embedding
of the circular supertube to the target space:
\beq
X^j=\sum_{m=-\infty}^\infty (\alpha_m^j e^{-im\theta}
+\bar{\alpha}_m^j e^{im \theta}), \;\;\;\; X^0=t,
\;\;\;\; Z \equiv X^9=z,
\label{xmu}
\eeq
where $j=1,2,\cdots,8$.
Let $2 \pi R$ as the 
circumference of the embedding circle in space-time.
Note that $\alpha_m^j ={\cal O} (R)$ and 
for the circular supertube,
$\alpha_1^1=\alpha_1^2=R/2$, others$=0$.
Though the vielbeins (and the induced metric)
will generically depends on $\theta$,
we can redefine the coordinate $\th$ $(0 \leq \th < 2\pi)$ 
such that
\beq
i \Dslash=\frac{1}{R} \sigma_1
\left( \frac{\de}{\de \theta} -i A_\th
\right) 
+\sigma_2 \frac{\de}{\de z},
\label{diract}
\eeq
i.e. $(d X^j)^2=R^2 d \theta^2$.
Now we consider the D2-brane
with (\ref{diract}) and (\ref{xmu}).
We take 
\beq
A_\th=Bz+\varphi, \;\;\;\;\; A_t=-E z, \;\; A_z=0,
\eeq
where $B$ and $\varphi$ is constant
in order to obtain the zero-modes easily
although $B$ can depend on $\th$ for 
assuring the supertube is a BPS state
(at least in the DBI approximation) \cite{MaNgTo}.

Then the configuration of the $\dz$-brane pairs for this D2-brane 
is given by
$T=u \Dslash$, 
$ \hat X^j$ and $ \hat Z$ 
corresponding to (\ref{xmu}) and $\hat A_t=-E Z$.
Because the Dirac operator has same form as (\ref{Dirac}),
we already know the zero-modes of the Dirac operator.
Thus we find 
\beqa
(\hat Z_0)_{mn} &=& \int d\theta dz \Psi_m^\dagger(\th,z) 
z \Psi_n (\th,z) 
=\frac{n-\varphi}{B} \delta_{m,n}, \nonumber \\
(\hat A_{t,0})_{mn} &=& -E \, (\hat Z_0)_{mn},
\label{vev2}
\eeqa
and what we should do is to compute the matrix elements
of $\hat X^j$ between the zero-modes.
We can easily compute it:
\beqa
(\hX^j_0)_{mn} &=& \int d\theta dz \Psi_m^\dagger(\th,z) 
\left(  \sum_{k=-\infty}^\infty (\alpha_k^j e^{-ik\theta}
+\bar{\alpha}_k^j e^{ik \theta}) \right)
\Psi_n (\th,z) \nonumber \\
&=&  \sum_{k=-\infty}^\infty 
\exp \left( -\frac{k^2}{4 BR} \right)
(\alpha_k^j \delta_{n,m-k}
+\bar{\alpha}_k^j \delta_{n,m+k} ).
\label{vev3}
\eeqa
Thus, on the Hilbert space $\{ \ket{m} \} $
we represent those as
\beqa
\hX^j_0 &=& \tilde{X}^j (\hat a^\dagger) \nonumber \\
\hat Z_0 &=& =\frac{\hat N-\varphi}{B} 
, \nonumber \\
\hat A_{t,0} &=& -E \frac{\hat N-\varphi}{B} 
,
\label{vev3a}
\eeqa
where 
\beq
\tilde{X}^j (e^{i\theta}) \equiv
 \sum_{k=-\infty}^\infty 
e^{ -\frac{k^2}{4 BR} } \;
(\alpha_k^j e^{-i k \theta} 
+\bar{\alpha}_k^j  e^{i k \theta}  ).
\label{xt}
\eeq
Note that if we take the base of the Hilbert space
as $e^{im\theta}=\ket{m}$,
we have
\beqa
\hX^j_0 = \tilde{X}^j (e^{i\theta}) , \;\;
\hat Z_0 = -i \frac{1}{B} \frac{\de}{\de \theta}-\frac{\varphi}{B}
=- \frac{1}{E} \hat A_{t,0}.
\label{vev4}
\eeqa
Here we introduce the coordinate eigen state
\beq
\ket{\theta}=\sum_m e^{im\theta} \ket{m },
\eeq
which satisfies $\hat a \ket{\theta}
= e^{i\theta} \ket{\theta}$
and $\hat a^\dagger \ket{\theta}
= e^{-i\theta} \ket{\theta}$.
Then, we have the following relations
\beq
\bra{\theta} A(a^\dagger) \ket{\theta'}
=2 \pi \delta(\theta- \theta') A(e^{i\theta})
\eeq
and
\beq
\Tr  A(a^\dagger)= 
\sum_m \bra{m} A(a^\dagger) \ket{m}
= \frac{1}{2\pi} \int_0^{2 \pi}  d\theta
\bra{\theta} A(a^\dagger) \ket{\theta}
= \delta(0) \int_0^{2 \pi} d\theta \; A(e^{i\theta}) .
\eeq
This implies that $\Tr 1 = 2 \pi \delta (0)$.
Because the density in the $z$ direction of the D0-branes 
is $B$,
we have $\Tr 1= \sum_m 1=BL$.
where $L$ is the IR cut off 
corresponds to the length of the $z$ direction.
Then we find
\beq
\Tr  A(a^\dagger) 
= \frac{BL}{2 \pi} \int_0^{2 \pi} d\theta \; A(e^{i\theta}) .
\eeq

The momentum given in the matrix model
is given by
\beq
\hat P_j=i [\hat Z_0, \hat X^j_0 ]=-i \frac{1}{B} 
\frac{\de \tilde{X}^j (\hat a^\dagger)}{\de \ln \hat a} 
=\left. \frac{1}{B} 
\frac{\de \tilde{X}^j (e^{i\theta})}{\de \theta} 
\right|_{e^{-i\theta} \rightarrow \hat a }. 
\eeq
Thus the total momentum density $\frac{1}{L} \Tr \hat P^j
= \int_0^{2 \pi} d\theta \frac{1}{2 \pi}
\frac{\de \tilde{X}^j (e^{i\theta})}{\de \theta}  $ vanishes.
The Hamiltonian of the matrix model for the supertube is \cite{BaLe}
\beq
{\cal H}=\frac{1}{2} \Tr \sum_{j=1}^8 \hat P_i^2
=L \frac{1}{B} \sum_{j=1}^8 
\frac{1}{2 \pi}  
\int_0^{2 \pi} d\theta \left( 
\frac{\de \tilde{X}^j (e^{i\theta})}{\de \theta} 
\right)^2,
\eeq
where we dropped the constant term which is proportional to $B$.
The fundamental string charge for the supertube 
in the matrix model is given by ${\cal H}/L$.

From the expression of the angular momentum
in the matrix model,
\beq
\hat L_{i j}=\hX^i_0 \hat P_j- \hX^j_0 \hat P_i,
\eeq
we find the total angular momentum density
in the matrix model as 
\beqa
\frac{1}{L} \Tr \hat L_{i j}
&=&
-\frac{1}{2 \pi}  
 \int_0^{2 \pi} d\theta \left( 
\tilde{X}^i (e^{i\theta})
\frac{\de \tilde{X}^j (e^{i\theta})}{\de \theta} 
-\tilde{X}^j (e^{i\theta})
\frac{\de \tilde{X}^i (e^{i\theta})}{\de \theta} 
\right) \\
&=& -\frac{1}{\pi}  \int_A d\tilde{X}^i
\wedge \; d\tilde{X}^j.
\eeqa
In the section \ref{D0D2}, 
we will compare these quantities with 
the corresponding ones for the supertube solutions
in the D2-brane picture.

We have assumed $B$ is a constant, however, it has been known that 
even if $B$ depends on $\theta$
the supertube is a BPS state in the DBI action.
In principle we can solve the zero-modes of the Dirac operator
for this case and obtain the D0-brane picture.
In this paper, however, we do not try to do it
because of a technical difficulty.
Instead, we will consider the supertube 
with the general cross section and a non-constant $B$ 
for the non-compact cross section
in the section \ref{Non-compact Cross Section}.

\subsection{Comparison between the DBI and the matrix model 
descriptions}
\label{D0D2}

In this subsection, we will compare the 
various quantities of the supertube solution
in the DBI description
and the matrix model description.
Here the DBI and the matrix model descriptions mean 
the D2-brane picture
using the DBI action which ignore the derivative corrections
and the D0-brane picture using 
the matrix model action which ignore the higher order terms,
respectively.
It it noted that the equivalence discussed in this paper
use the boundary state or boundary string field theory,
thus the configuration in the D0-branes 
is exactly same as the configuration in the D2-brane.
However, there will be differences in the two descriptions
if we approximate it in the different ways.

First, we consider the D0-brane charge.
We have seen that the D0-brane charge density along $z$ axis 
in the matrix model is $B$, which is not corrected by
the higher order terms in the matrix model
because it is essentially the size of the matrices.
In the DBI picture, 
we can commutate the D0-brane charge from the Chern-Simons term as
$\frac{1}{L} \frac{1}{2 \pi} \int d\theta \int dz B=B$.
Actually, 
these two computations should give same values
since the Chern-Simons term may not be corrected by
the derivative corrections.

However, the fundamental string charges will be 
different in two pictures.
In the DBI description it is computed as \cite{MaNgTo}
\beq
q_{F1}=\frac{1}{L} \frac{1}{2 \pi} \int d\theta \int dz 
\frac{1}{B} \sum_{j=1}^8 (\frac{\de X^j}{\de \theta})^2
=\frac{1}{2 \pi B}  
\int d\theta \sum_{j=1}^8 (\frac{\de X^j}{\de \theta})^2
=\frac{R^2}{B},
\label{fsc}
\eeq
for a constant $B$.
On the other hand, the one computed in the matrix model
is same form as (\ref{fsc}), however, $X^j$ should be  
replaced by $\tilde X^j$ which was defined by (\ref{xt}).
Thus, for example, the fundamental string charge in the matrix model 
for the circular supertube
is $q_{F1}=e^{-\frac{1}{2BR}} R^2/B$.
For the angular momentum, there is a similar difference
between the two descriptions.
These differences will be originated from 
the different approximations and will be absent 
if we includes all higher order and higher derivative corrections.
We note that 
if the dimensionless parameter $BR$ 
is very large $BR \gg 1$, the differences become very small
because $\tilde X^j \sim X^j$ for $BR \gg 1$.\footnote{
For $BR \gg 1$,
$\exp (-\frac{1}{2 BR} ) \sim 1- \frac{1}{2 BR} 
+ {\cal O} (\frac{1}{B^2 R^2}) $
will be corrections to the DBI actions or the matrix model, but 
it is not $\alpha'$ correction, but like $1/N$ corrections.}

The higher order corrections for the matrix model action 
are expected to contain the commutators.
If we assume the D0-brane action is given by the
T-dual of the usual D$p$-brane actions,
the corrections to the matrix model actions 
includes the terms corresponding to the DBI action,
i,e, $[\hX,\hX]^n \sim {\cal O} (\frac{R^n}{B^n}) $,
and terms corresponding to the higher derivative terms,
i.e. $[\hX, [\hX, \cdots, [\hX,\hX],],\cdots] 
\sim {\cal O} (\frac{1}{B^n})$.
Thus the matrix model action may be reliable 
for $|B/R| \gg 1$ and $|B| \gg 1$.
On the other hand, the DBI action may be reliable for 
$R \gg 1$.
Thus $|BR| \gg 1$ is necessary for
reliability of the both descriptions.
We will discuss the ambiguity of the matrix model 
in subsection \ref{comments} again.

\section{Some Generalizations}
\label{gene}

\subsection{Supertubes with non-compact cross section}
\label{Non-compact Cross Section}

Here we consider the flat super''tube''.
First, we consider
the flat D2-brane with 
$(2 \pi \alpha') F=E dt \wedge dy +B dy \wedge dx$ 
in the flat spacetime.
(Here $x$ and $y$ corresponds to the $\theta$ and $z$, respectively,
in the previous section.)
It is clear that this is a solution of the equations of motion
for any constant $B,E$. Especially, For $E=0$ this is
the usual noncommutative D2-brane \cite{CoDoSc, DoHu, SeWi}.
Below we will assume $B>0$.
The Dirac operator in a particular gauge is
\beq
i \Dslash= \sigma_x \left( \frac{\de}{\de x} -i B y \right)
+\sigma_y  \frac{\de}{\de y}.
\eeq
In this case the zero-modes of the Dirac operator
was already given in \cite{Te2,El} and 
the equivalent D0-brane picture was obtained. %\footnote{
%We will study some deformations of it 
%and a relation to the Seiberg-Witten map in the Appendix.} 
However, here we will use different basis of the zero-modes
parametrised by a real number $\zeta$:
\beq
\Psi_\zeta (x,y) =
\left( \frac{|B|^3}{4 \pi^3} \right)^\frac{1}{4} 
\exp \left( i B y (x-\zeta) 
-\frac{|B|}{2} \left( x-\zeta \right)^2 
\right) 
\left(
\begin{array}{cc} 0 \\ 1  \end{array} \right).
\eeq
where $\Psi_\zeta$ is normalized such that,
\beq
\int dx dy \Psi_\zeta^\dagger \Psi_{\zeta'} 
=\delta (\zeta-\zeta').
\eeq
Then the matrix elements of 
$\hat X=x,\hat Y=y$ 
and $\hat A_t=-E y$
between the zero-modes
are computed as
\beqa
(\hat X_0)_{\zeta \zeta'} &=& \int dx dy \Psi_\zeta^\dagger
x \Psi_{\zeta'} =\zeta \delta (\zeta-\zeta'), \;\; \nonumber \\
(\hat Y_0)_{\zeta \zeta'} &=& \int dx dy \Psi_\zeta^\dagger
y \Psi_{\zeta'} = \frac{1}{i B}
\frac{\de }{\de \zeta'} \delta (\zeta-\zeta'), \;\; \nonumber \\
(\hat A_{t,0})_{\zeta \zeta'} &=& -E \, (\hat Y_0)_{\zeta \zeta'},
\label{vevnonc}
\eeqa
which can be represented in the Hilbert space spanned
by $\ket{x}$, as $\hat X_0=\hat x$ and $\hat Y_0=\hat y$
where $\hat{x} \ket{x}=x \ket{x}$ and $[\hat x, \hat y]= i/B$.
This means that the D2-brane is equivalent to 
the infinitely many D0-branes with (\ref{vevnonc}) as expected.
%In the boundary state formalism, this was shown in \cite{Is}.
We can also compute
\beq
\int dx dy \, \Psi_\zeta^\dagger \;
e^{i \alpha x+i \beta y}
\; \Psi_{\zeta'} 
=e^{-\frac{1}{4B} (\alpha^2+\beta^2+2i \alpha \beta) }
e^{i\zeta' \alpha} 
\delta \left( \zeta-\zeta'+\frac{\beta}{B} \right).
\label{gv1}
\eeq
This can be represented on $\ket{x}$ as
\beq
e^{-\frac{1}{4B} (\alpha^2+\beta^2+2i \alpha \beta) }
 e^{i \beta \hat{y} } e^{i \alpha \hat{x}} 
= e^{-\frac{1}{4B} (\alpha^2+\beta^2) } 
e^{i \alpha \hat{x} +i \beta \hat{y} }
= e^{i \xi \hat{\eta}} e^{i \bar{\xi} \hat{\eta}^\dagger },
\label{gv2}
\eeq
where we have defined
\beq
\hat{\eta} \equiv \sqrt{\frac{B}{2} } (\hat{x}+i \hat{y} ), 
\;\;\; \xi \equiv  \sqrt{\frac{1}{2B}} (\alpha -i \beta).
\label{aa}
\eeq
Note that $\hat \eta$ is a lowering operator
\beq
[\hat{\eta}, \hat{\eta}^\dagger ]=1,
\eeq
and (\ref{gv2}) is anti-normal ordered.
This result means that 
if the D2-brane has the nontrivial transverse coordinates
$X^i=X^i(x,y)$ $(i=2,3, \cdots, 8)$, 
the D0-branes have the transverse coordinates
$\hat X^i_0=\left[ X^i(\hat x, \hat y) \right]_A$
where $[\dots ]_A$ means the anti-normal ordering.

Now let us consider the non-compact supertube 
with an arbitrary magnetic field.
It is the D2-brane with 
$F=E dt \wedge dy +B(x) dy \wedge dx$ 
and $X^i=X^i(x)$ $(i=2,3, \cdots, 8)$ 
in the flat spacetime.
Interestingly, for $E=\pm 1$ it was shown in \cite{KrMyPeWi}
that this is indeed a BPS state for all orders in the $\alpha'$ expansion.
The coordinate $x$ can be chosen such that
the Dirac operator becomes
\beq
i \Dslash= \sigma_x \left( \frac{\de}{\de x} -i B y \right)
+\sigma_y  \left( \frac{\de}{\de y} -i A_y (x) \right).
\eeq
where we took a particular gauge for the magnetic field 
$B(x)$ such that
$B(x)=B-\de_x A_y(x)$ where $B$ is the constant part 
of $B(x)$
and, $A_0=-E y$, $X^i=X^i(x)$ $(i=3,4, \cdots, 9)$.
Then the zero-modes are
\beq
\Psi_\zeta (x,y) =
C_\zeta
\exp \left( i B y (x-\zeta) 
-\frac{|B|}{2} \left( x-\zeta \right)^2 
+ F(x)
\right) 
\left(
\begin{array}{cc} 0 \\ 1  \end{array} \right).
\eeq
where 
\beq
F(x) \equiv \int^x_0 dx' A_y(x'),
\eeq
and $C_\zeta$ is the normalization constant 
to be determined by
$ \int dx dy \Psi_\zeta^\dagger \Psi_{\zeta'} 
=\delta (\zeta-\zeta')$.
The matrix elements are given by
\beqa
(\hat X_0)_{\zeta \zeta'} &=& \int dx dy \Psi_\zeta^\dagger
x \Psi_{\zeta'} =h(\zeta) \delta (\zeta-\zeta') 
, \;\; \nonumber \\
(\hat Y_0)_{\zeta \zeta'} &=& \int dx dy \Psi_\zeta^\dagger
y \Psi_{\zeta'} = \frac{1}{i B}
\frac{\de }{\de \zeta'} \delta (\zeta-\zeta'), \;\; 
(\hat A_{t,0})_{\zeta \zeta'} = -E \, (\hat Y_0)_{\zeta \zeta'},
\;\; \nonumber \\
(\hat X^i_0)_{\zeta \zeta'} &=& \int dx dy \Psi_\zeta^\dagger
X^i(x) \Psi_{\zeta'} =h^i (\zeta) \delta (\zeta-\zeta') 
\label{vevnonc2}
\eeqa
where 
\beq
h(\zeta) \equiv  \zeta +
\frac{ \int dx x e^{-|B| x^2 +2 F(x+\zeta) } }
{ \int dx e^{-|B| x^2 +2 F(x+\zeta) } }, \;\;\;\;
h^i(\zeta) \equiv 
\frac{ \int dx X^i(x) e^{-|B| (x-\zeta)^2 +2 F(x) } }
{ \int dx e^{-|B| (x-\zeta)^2 +2 F(x) } }.
\eeq
These can be represented as $\hat X_0=h(\hat x) $, $\hat Y_0=\hat y$
and $\hat X^i_0=h^i(\hat x)$
and indeed satisfy the BPS equations of the matrix model.\footnote{
For the pure gauge $F(x)=a x +c$, we have $\hX_0=\hat x -a/B$.
We can redefine $\hat x' =\hat x-a/B$ which does not change the 
commutation relation. For $F(x)=-b x^2 /2$, 
we have $\hX_0=\frac{B}{B+b} \hat x$, which
means the shift of $B$ to $B+b$ 
because $[\hX_0, \hat Y_0]=i/(B+b)$.
This is consistent with the fact that $F(x)=-b x^2 /2$
is the shift of $B$. 
}

In the Appendix, we will consider the inverse of 
the transformation,
i.e. 
the D2-brane supertube from the supertube in the matrix model 
using infinitely many D2-brane-anti-D2-brane pairs.

\subsection{Superfunnel type solutions}

In this subsection, instead of 
$A_t=-E y$, 
we take $A_t = -E Z$ where $E= \pm 1$.
Then for 
$\triangle z(x,y)=0$ and $x^i=0 \;\;\; (i=3,\cdots,8)$, 
the D2-brane is the noncommutative BIon, or 
the superfunnel. 
The zero-modes of the tachyon are same as in the previous subsection.
Thus using (\ref{gv1}) and (\ref{gv2}),
the D0-branes have the transverse coordinates
$\hat Z_0=\left[ z(\hat x, \hat y) \right]_A$,
where $[\dots ]_A$ means the anti-normal ordering.
This satisfies the equations of motion and the Gauss law 
of the matrix model
(\ref{lap}) because the derivation keeps the anti-normal ordering.

The D$2p$-brane with
$F=E dt \wedge d z+ b_a dy^a \wedge dx^a \;\;\; (a=1,\cdots, p) $ 
where $E= \pm 1$,
$\triangle \, z(x_a,y_a)=0$ 
and $x^i=0 \;\;\; (i=2p+1,\cdots,8)$, 
is the noncommutative BIon, 
which is the solution of the equation of motion of 
the $N=4$ supersymmetric Yang-Mills actions.
The equivalent D0-branes are easily obtained as before,
and we have $\hat Z_0=\left[ z(\hat x^a, \hat y^a) \right]_A$
although the solution of the equations of motion 
of the matrix model
\beq
\sum_{a=1}^p \frac{1}{b_a^2} 
\left( \frac{\de^2 }{\de^2 \hX^{a}} 
+\frac{\de^2 }{\de^2 \hat Y^{a}} \right) \hat Z(\hX, \hat Y)=0,
\label{eq4}
\eeq
is different from $\triangle \, z(x_a,y_a)=0$
except when all $b_a$s are same.
However, in this case, we do not know 
if the equations of motion should be corrected beyond
the Yang-Mills or the matrix model approximation.
Note that the former is valid for $|b_a| \ll 1$ and 
$r \equiv \sqrt{(x^a)^2+(y^a)^2 } \gg 1 $
and the latter is valid for $|b_a|  \gg 1$ and $r \gg 1$.
Moreover, it is possible that
two configurations can be matched
by some field redefinition.
Indeed, by the field redefinition of the theory on the D$(2p)$-brane,
we can use the noncommutative Yang-Mills actions
with $\theta=1/B, \; \Phi=-B$ and $G=-B\frac{1}{g} B$,
i.e. $S \sim \sqrt{G} G^{-1} ((\hat F_*-B))^2$ which 
was shown to be equivalent to the matrix model action \cite{Seiberg}.
Then, the equations of motion of the noncommutative Yang-Mills 
action is same as (\ref{eq4}) because of the $G^{-1} \sim 1/ |b^a|$ factor.
This is natural since
the string world renormalization conditions for 
the D2-brane and the D0-branes will be same for this choice
of the fields
as seen from \cite{Seiberg}.\footnote{
In \cite{Seiberg} the Weyl ordered product and the Moyal star product
are used. However, any ordering can be used in the matrix model description
of the noncommutative gauge theory,
especially the anti-normal ordered product.
Actually, it can be shown as in \cite{Seiberg} that
the equivalence between the matrix model with the anti-normal ordering
and the noncommutative gauge theory
of $\Phi=-B$ with the star product of the anti-normal ordering.
Note that the choice of the ordering in the matrix model 
does not change the operator, for example $\hat A_i(\hat x)$, though
we should redefine the noncommutative gauge field
in the noncommutative gauge theory side.
}

\subsection{Comments on the ambiguity of the effective action of 
the D0-branes}
\label{comments}

In the previous sections, 
we take the effective action of 
the D0-branes as the matrix models with 
the higher order corrections derived from the 
T-dual, i.e. the dimensional reduction of the DBI actions.\footnote{
For the non-Abelian case, there are ordering ambiguities,
which is, however, can be fixed by the string world sheet computation
in principle.}
However, the effective actions of the D-branes
have the field redefinition ambiguity, which is interpreted 
as the regularization (or, more precisely, the finite renormalization) 
of the string world sheet action,
in general.
Therefore, in order to compare the physical quantities
in the D0-brane and D$(2p)$-brane pictures,
we should use the effective actions obtained from
the string world sheet computations with 
the same regularization, or more precisely the same 
renormalization condition,
even though the two boundary states
in the D0-brane and D$(2p)$-brane pictures are equivalent
and every quantities should be same in principle.\footnote{
We would have to choose some regularization such that
the reduction to the zero-modes of the Dirac operator 
are valid after the regularization 
and the equivalence between the path-integral and the 
operator formalism used in \cite{AsSuTe3} to show the equivalence
between D-brane systems is still valid.  }

The effective action of the D0-branes given by the T-dual 
are based on the 
very special choice of the regularization of 
the world sheet although it is a natural choice.
For the flat noncommutative D$(2p)$-brane solution in the D0-branes,
this choice corresponds to the 
the noncommutative parameter $\theta=1/B$ and $\Phi=-B$ 
as seen from the results of \cite{Seiberg}.
However, if we make the following field redefinition,
$\hX^i \ra \hX^i+\beta^{kl} [\hX^k,\hX^l] \hX^i$,
we have a different noncommutative parameter 
$\theta \equiv [\hX^i, \hX^j]=(1/B)_{ij} \ra ((1+\beta/B)^2/B)_{ij}$.
Note that the $U(N)$ symmetry of the matrix model
should contain the noncommutative $U(1)$ gauge symmetry 
of the D$(2p)$-brane
with any noncommutative parameter $\theta$, including
usual commutative gauge symmetry
since there are the Seiberg-Witten maps in the D$(2p)$-brane
effective actions.

From this observation of the noncommutative D$(2p)$-brane,
we conclude that
we have to refine the fields, say $\hat X^i$, 
to compare the physical quantities computed using 
the usual commutative DBI action of the D$2p$-brane and 
the matrix model of the D0-branes derived from the 
the dimensional reduction of the DBI action.

We also note that
there should be some differences in the renormalization conditions
between  
the computations directly using the result of \cite{Is} and 
ours using the $\dz$-brane pairs.

%\section{Conclusions}
%\label{concl}
%In this paper, 
%we have studied 
%
%For the noncommutative BIon case, 
%*********************************************

\subsection{Comments for the Fuzzy $S^2$ case}
\label{fuzzy}

Finally, we will briefly comment 
on the fuzzy $S^2$ case comparing to the supertube case.
In \cite{Te2}, the D2-brane on the fuzzy $S^2$ was
considered using the tachyon condensation in the same way 
in this paper.
In the fuzzy $S^2$ case, the DBI action would be 
reliable for $R \gg 1$ (or $R \gg l_s$ if we recover the 
$\alpha'$ dependence), but
the matrix model would be reliable for $1/M_{D0} \gg 1$ 
(or $1/M_{D0} \gg l_s$) 
where $1/M_{D0} \equiv N l_s^2 /R$.\footnote{
Here $R$ is the radius of the $S^2$ and 
$N$ is the number of the D0-branes.}
Moreover, $N=R M_{D0}$ (which is a dimensionless parameter)
controls the similarity of two descriptions,
i.e. in the $N \rightarrow \infty$ limit the two descriptions 
are same.
We have seen that 
the parameter $BR$ for the supertube plays a similar role
as $N$ of the fuzzy $S^2$ case. 
Actually, $BR$ is approximately 
the number of the 
D0-branes in a sphere of radius $R$ centered at 
a point on $Z$ axis.

Here we note that 
$N=1$ and $N=0$ are special.
The matrix model is 
always reliable for these cases because $[\hX^i,\hX^j]=0$.
This $N=1$ case is like the supertube with 
$BR =0$ in which the effective radius is infinitely small.
%however, for the $N=1$ case D0-brane picture 
%is good for arbitrary $R$.
(This $N=1$ case is also similar to the tachyon condensation
of two non BPS D0-branes to a non BPS D0-brane in \cite{Te2}.)
Furthermore, if we take $R \gg 1$ for these,
both the matrix model and the DBI descriptions seem good
although these two descriptions are apparently different.
%This is because the fuzzy $S^2$ configuration
%is an off-shell configuration.
However, the boundary state with the path-integral 
over the curved world volume
used in \cite{Te2} 
might be inappropriate for these cases.
Indeed, if we take very large, but finite $u$,
the D2-brane is localized on $S^2$ with the radius $R$.\footnote{
Here the localized brane means localized in the usual sense,
for example, the supertube with $R \gg 1$ and $|B| \ll 1$.
In \cite{Te2} the noncommutativity means 
the noncommutativity in the D0-picture, i.e. $[\hX^i, \hX^j]
\neq 0$, and
the localized branes means
the branes which is commutative in the D0-picture.
However, the localized brane in the usual sense 
can be ``non-localized'' in the D0-picture.
Thus the term ``localized'' in \cite{Te2} is not appropriate
in the usual sense.}

In order to see how the fuzzy $S^2$ 
in an on-shell configuration behaves,
let us consider the $N$ monopoles in the D3-branes \cite{HaTe3}.
Here $N<0$ case corresponds to $|N|$ anti-monopoles.
The radius of the fuzzy sphere 
is $R \sim |N| / (\xi \pm a/2)$ where $\xi$ is 
the coordinate along the D1-branes and D3-branes are at 
$\xi \pm a/2$.
The D1-brane picture is good for $ |N| /R \gg 1 $,
on the other hand,
the D3-brane picture is good for $R  \gg 1$ \cite{ChWe}. \footnote{
For the $k$ instantons, the D4-brane picture is 
valid if a scale of the instantons 
$\rho$ is very large. 
On the other hand, D0-brane picture is 
valid if the mass scale of the D0-branes $M$ is large
and there is a relation $\rho M \sim |k|$. }
For $|N|=1$, the D1-brane can not form the fuzzy sphere.
Nevertheless the D1-brane picture will be valid 
for $a \gg 1$ and $|\xi \pm a/2| \gg 1 $ which means $R \ll 1$.
Near the boundary $\xi \sim \pm a/2$, 
the D1-brane picture for $|N|=1$ 
is singular, but the D3-brane picture is valid.
(More precisely, for $|N|=1$
the singularities at the boundaries are worse than $|N|>1$ case. )
Two pictures should be smoothly connected.
In this case we took $u \ra \infty$ limit, but 
the D3-brane picture 
is valid near $|\xi \pm a/2| \sim 0 $
because the D1-brane picture should be corrected near the singularity
$|\xi \pm a/2| \sim 0 $.
Thus it is interesting to see if the difference of the radii
of the supertubes with the compact cross section in the two pictures
is an artifact or not.
We note that in the monopole case, 
although the embedding into the flat spacetime is nontrivial,
we have not used the path-integral over the curved world volume.

%Thus the D1 picture is bad 
%near the boundaries where the D3 picture is good.

%However, to show the equivalence between the D0 and D2 pictures,
%we need to expand by $\Phi^i \sim R$.
%Thus for $R \gg 1$ the equivalence would not hold and
%the D2-picture would be different from the D0-picture
%in these cases. 

\vskip6mm
\noindent
{\bf Acknowledgements}

\vskip2mm
The author would like to thank K.~Hashimoto, I.~Kishimoto, K.~Murakami
and T.~Takayanagi
for useful discussions.
\\

\noindent
%{\bf Note added}: 

\appendix
\setcounter{equation}{0}

\section{The inverse transformation}
\label{invers}

In \cite{HaTe3, HaTe4},
both of 
the ADHM(N) construction of instanton (monopole)
and the inverse ADHM(N) construction were obtained 
using the tachyon condensation.
In this appendix we will find 
the inverse transformation for the supertube with the non-compact cross section.
Here the inverse transformation for the supertube
means that the construction of the supertube 
from the infinitely many D2-brane and anti-D2-brane pairs
instead of the $\dz$-branes.

First, let us remember that using the decent relations \cite{Senreview}
a D0-brane at $x^1=a^1, x^2=a^2$ can be constructed from
a pair of the D2-brane and the anti-D2-brane
by giving $T=u \sigma_1 (x^1-a^1)+\sigma_2 (x^2-a^2))$
and $A_1=A_2=0$
\cite{KrLa, TaTeUe}.
Thus, we can obtain 
the infinitely many D0-branes with $\hat X^i=\hat x^i$
where $[\hat x^1,\hat x^2]=i/B$ 
from  infinitely many pairs of the D2-brane and the anti-D2-brane
with
\beq
T(x^1,x^2)=u (\sigma_1 (x^1-\hat x^1)+\sigma_2 (x^2-\hat x^2))
\eeq
and $A_1=A_2=0$.
Furthermore, $\hat A_t=E \hat x^2$ is induced in the D0-brane picture
by setting $A_t= E \hat x^2$ for the pairs of 
the D2-brane and the anti-D2-brane.

Then, to obtain the D2-brane picture,
we should solve 
\beq
\Dslash \psi(z)=\matt[0,\bar z-\hat z^\dagger,z-\hat z,0]
\left(
\begin{array}{cc} \psi_1 (z) \\ \psi_2 (z)  \end{array} \right)=0,
\eeq
where $z=x^1+ix^2$ and $\hat z=\hat x^1+i \hat x^2$.
There is only one normalizable zero-mode of the tachyon 
which corresponds to the surviving 
D2-brane after the tachyon condensation:
\beq
\psi_1(z)=\ket{z}, \;\; \psi_2=0,
\eeq
where $\ket{z}$ is the normalized coherent state 
$\hat z \ket{z}=z \ket{z}$.
This zero-mode depends on $z$, thus 
we have \cite{HaTe3}
\beq
A_z=2 i \psi^\dagger D_{\bar z} \psi=-i \frac{B}{2} z,
\eeq
where $A_z=A_1+i A_2$ and 
$D_z=\frac{1}{2}(\frac{\de}{\de x^1}+\frac{\de}{\de x^2})$.
Note that
it can be considered as the Berry phase 
although it is not an approximation, but an exact 
result in our $u \ra \infty$ limit.
Thus we have the D2-brane with $F_{12}=-B$.
For the time component of the gauge field
we have
\beq
A_t=\psi^\dagger E \hat x^2 \psi =E x^2. 
\eeq
Thus we have shown the equivalence by giving the 
map from the D0-brane picture to the D2-brane picture.

We can also consider
the non-compact supertube in the D0-brane picture
with the nontrivial transverse coordinate
\beq
\hat X^i= \int d\alpha^i \int d\beta^i \;
C(\alpha^i,\beta^i)  \; e^{i \bar{\xi^i} \hat{\eta}^\dagger } 
e^{i \xi^i \hat{\eta}},
\label{normal}
\eeq
where $\xi^i =  \sqrt{\frac{1}{2B}} (\alpha^i -i \beta^i)$
and $\hat{\eta} \equiv \sqrt{\frac{B}{2} } (\hat{x}^1+i \hat{x}^2 )$ .
Then in the D2-brane picture it becomes
\beq
X^i(x^1,x^2)=\int d\alpha^i \int d\beta^i \;
C(\alpha^i,\beta^i)  \;  e^{i \alpha^i x^1+i \beta^i x^2}.
\eeq
Here we note that (\ref{normal}) is normal ordered
instead of anti-normal ordered used in (\ref{gv2}).
This difference could be originated from the possible 
field redefinition.

%\section{Boundary state}
%\label{bsr}

\newpage

%%%%%%%%%% References %%%%%%%%%%%%%%%%%%%%%%%%%
\newcommand{\J}[4]{{\sl #1} {\bf #2} (#3) #4}
\newcommand{\andJ}[3]{{\bf #1} (#2) #3}
\newcommand{\AP}{Ann.\ Phys.\ (N.Y.)}
\newcommand{\MPL}{Mod.\ Phys.\ Lett.}
\newcommand{\NP}{Nucl.\ Phys.}
\newcommand{\PL}{Phys.\ Lett.}
\newcommand{\PR}{ Phys.\ Rev.}
\newcommand{\PRL}{Phys.\ Rev.\ Lett.}
\newcommand{\PTP}{Prog.\ Theor.\ Phys.}
\newcommand{\hep}[1]{{\tt hep-th/{#1}}}
%%%%%%%%%%%%%%%%%%%%%%%%%%%%%%%%%%%%%%%%%%%%%%%


\begin{thebibliography}{99}
\baselineskip=13pt

\bibitem{Wi}
  E.~Witten,
  ``Small Instantons in String Theory,''
  Nucl.\ Phys.\ B {\bf 460} (1996) 541,
  hep-th/9511030.
  %%CITATION = HEP-TH 9511030;%%


\bibitem{Do}
  M.~R.~Douglas,
  ``Branes within branes,''
  hep-th/9512077.
  %%CITATION = HEP-TH 9512077;%%

\bibitem{Is}
N. Ishibashi,
``$p$-branes from $(p-2)$-branes
in the Bosonic String Theory,''
Nucl. Phys. {\bf B539} (1999) 107,
hep-th/9804163.


\bibitem{HiNoSu}
  Y.~Hikida, M.~Nozaki and Y.~Sugawara,
  ``Formation of spherical D2-brane from multiple D0-branes,''
  Nucl.\ Phys.\ B {\bf 617} (2001) 117
  , hep-th/0101211.
  %%CITATION = HEP-TH 0101211;%%

\bibitem{Te2}
  S.~Terashima,
  ``Noncommutativity and tachyon condensation,''
  JHEP {\bf 0510} (2005) 043
  , hep-th/0505184.
  %%CITATION = HEP-TH 0505184;%%



\bibitem{HaTe3}
  K.~Hashimoto and S.~Terashima,
  ``Stringy derivation of Nahm construction of monopoles,''
  JHEP {\bf 0509} (2005) 055
  , hep-th/0507078.
  %%CITATION = HEP-TH 0507078;%%


\bibitem{HaTe4}
  K.~Hashimoto and S.~Terashima,
  ``ADHM is tachyon condensation,''
  JHEP {\bf 0602} (2006) 018
  , hep-th/0511297.
  %%CITATION = HEP-TH 0511297;%%



\bibitem{MaTo}
  D.~Mateos and P.~K.~Townsend,
  ``Supertubes,''
  Phys.\ Rev.\ Lett.\  {\bf 87} (2001) 011602
  , hep-th/0103030.
  %%CITATION = HEP-TH 0103030;%%


\bibitem{BaLe}
  D.~Bak and K.~M.~Lee,
  ``Noncommutative supersymmetric tubes,''
  Phys.\ Lett.\ B {\bf 509} (2001) 168
  , hep-th/0103148.
  %%CITATION = HEP-TH 0103148;%%


\bibitem{EmMaTo}
  R.~Emparan, D.~Mateos and P.~K.~Townsend,
  ``Supergravity supertubes,''
  JHEP {\bf 0107} (2001) 011
  , hep-th/0106012.
  %%CITATION = HEP-TH 0106012;%%


\bibitem{BaKa}
  D.~Bak and A.~Karch,
  ``Supersymmetric brane-antibrane configurations,''
  Nucl.\ Phys.\ B {\bf 626} (2002) 165
  , hep-th/0110039.
  %%CITATION = HEP-TH 0110039;%%


\bibitem{MaNgTo}
  D.~Mateos, S.~Ng and P.~K.~Townsend,
  ``Tachyons, supertubes and brane/anti-brane systems,''
  JHEP {\bf 0203} (2002) 016
  , hep-th/0112054.
  %%CITATION = HEP-TH 0112054;%%


\bibitem{BaOh}
  D.~s.~Bak and N.~Ohta,
  ``Supersymmetric D2 anti-D2 strings,''
  Phys.\ Lett.\ B {\bf 527} (2002) 131,
  hep-th/0112034.
  %%CITATION = HEP-TH 0112034;%%


\bibitem{KrMyPeWi}
  M.~Kruczenski, R.~C.~Myers, A.~W.~Peet and D.~J.~Winters,
  ``Aspects of supertubes,''
  JHEP {\bf 0205} (2002) 017
  , hep-th/0204103.
  %%CITATION = HEP-TH 0204103;%%

\bibitem{Ma}
  S.~D.~Mathur,
  ``The fuzzball proposal for black holes: An elementary review,''
  Fortsch.\ Phys.\  {\bf 53} (2005) 793
  , hep-th/0502050.
  %%CITATION = HEP-TH 0502050;%%

\bibitem{BaOhSh}
  D.~Bak, N.~Ohta and M.~M.~Sheikh-Jabbari,
  ``Supersymmetric brane anti-brane systems: Matrix model description,
  stability and decoupling limits,''
  JHEP {\bf 0209} (2002) 048,
  hep-th/0205265.
  %%CITATION = HEP-TH 0205265;%%

\bibitem{BaOhTo}
  D.~Bak, N.~Ohta and P.~K.~Townsend,
  ``The D2 susy zoo,''
  hep-th/0612101.
  %%CITATION = HEP-TH 0612101;%%








\bibitem{Ta}
  H.~Takayanagi,
  ``Boundary states for supertubes in flat spacetime and Goedel universe,''
  JHEP {\bf 0312} (2003) 011
  , hep-th/0309135.
  %%CITATION = HEP-TH 0309135;%%



\bibitem{Sen}
  A.~Sen,
  ``Tachyon condensation on the brane antibrane system,''
  JHEP {\bf 9808} (1998) 012,
  hep-th/9805170;
  %%CITATION = HEP-TH 9805170;%%
%\bibitem{Sen:1999mh}
%  A.~Sen,
  ``Descent relations among bosonic D-branes,''
  Int.\ J.\ Mod.\ Phys.\ A {\bf 14} (1999) 4061,
  hep-th/9902105;
  %%CITATION = HEP-TH 9902105;%%
%\bibitem{Sen:1999mg}
%  A.~Sen,
  ``Non-BPS states and branes in string theory,''
  hep-th/9904207;
  %%CITATION = HEP-TH 9904207;%%
%\bibitem{Sen:1999xm}
%  A.~Sen,
  ``Universality of the tachyon potential,''
  JHEP {\bf 9912} (1999) 027,
  hep-th/9911116.
  %%CITATION = HEP-TH 9911116;%%


\bibitem{Senreview}
A.~Sen, ``Tachyon dynamics in open string theory,''
hep-th/0410103.
%%CITATION = HEP-TH 0410103;%%




\bibitem{AsSuTe1}
%\bibitem{Asakawa:2001vm}
  T.~Asakawa, S.~Sugimoto and S.~Terashima,
  ``D-branes, matrix theory and K-homology,''
  JHEP {\bf 0203} (2002) 034
  , hep-th/0108085;
  %%CITATION = HEP-TH 0108085;%%




\bibitem{KMM2}
D.~Kutasov, M.~Marino and G.~W.~Moore, ``Remarks on tachyon
condensation in superstring field theory,'' hep-th/0010108.
%%CITATION = HEP-TH 0010108;%%

\bibitem{KrLa}
P.~Kraus and F.~Larsen, ``Boundary string field theory of the
DD-bar system,'' Phys.\ Rev.\ D {\bf 63} (2001) 106004
, hep-th/0012198.
%%CITATION = HEP-TH 0012198;%%


\bibitem{TaTeUe}
T.~Takayanagi, S.~Terashima and T.~Uesugi, ``Brane-antibrane
action from boundary string field theory,'' JHEP {\bf 0103} (2001)
019, hep-th/0012210.




\bibitem{El}
  I.~Ellwood,
  ``Relating branes and matrices,'' JHEP {\bf 0508} (2005) 078,
  hep-th/0501086.
  %%CITATION = HEP-TH 0501086;%%




\bibitem{Te}
S.~Terashima, ``A construction of commutative D-branes from lower
dimensional non-BPS D-branes,'' JHEP {\bf 0105}, 059 (2001)
, hep-th/0101087.
%%CITATION = HEP-TH 0101087;%%




\bibitem{AsSuTe2}
%\bibitem{Asakawa:2002nv}
  T.~Asakawa, S.~Sugimoto and S.~Terashima,
  ``D-branes and KK-theory in type I string theory,''
  JHEP {\bf 0205} (2002) 007
  , hep-th/0202165;
  %%CITATION = HEP-TH 0202165;%%


\bibitem{AsSuTe3}
  T.~Asakawa, S.~Sugimoto and S.~Terashima,
  ``Exact description of D-branes via tachyon condensation,''
  JHEP {\bf 0302} (2003) 011
  , hep-th/0212188;
%\bibitem{Asakawa:2003ax}
%  T.~Asakawa, S.~Sugimoto and S.~Terashima,
  ``Exact description of D-branes in K-matrix theory,''
  Prog.\ Theor.\ Phys.\ Suppl.\  {\bf 152} (2004) 93
  , hep-th/0305006;
  %%CITATION = HEP-TH 0305006;%%



\bibitem{BFSS}
T.~Banks, W.~Fischler, S.~H.~Shenker and L.~Susskind, ``M theory
as a matrix model: A conjecture,'' Phys.\ Rev.\ D {\bf 55} (1997)
5112 , hep-th/9610043.
%%CITATION = HEP-TH 9610043;%%


\bibitem{Hy}
  Y.~Hyakutake,
  ``Fuzzy BIon,''
  Phys.\ Rev.\ D {\bf 68} (2003) 046003
  , hep-th/0305019.
  %%CITATION = HEP-TH 0305019;%%



\bibitem{CaMa}
  C.~G.~.~Callan and J.~M.~Maldacena,
  ``Brane dynamics from the Born-Infeld action,''
  Nucl.\ Phys.\ B {\bf 513} (1998) 198
  , hep-th/9708147.
  %%CITATION = HEP-TH 9708147;%%

\bibitem{Gi}
  G.~W.~Gibbons,
  ``Born-Infeld particles and Dirichlet p-branes,''
  Nucl.\ Phys.\ B {\bf 514} (1998) 603,
  hep-th/9709027.
  %%CITATION = HEP-TH 9709027;%%


\bibitem{Seiberg}
  N.~Seiberg,
  ``A note on background independence in noncommutative gauge theories,  matrix
  %model and tachyon condensation,''
  JHEP {\bf 0009} (2000) 003
  , hep-th/0008013.
  %%CITATION = HEP-TH 0008013;%%








\bibitem{CoDoSc}
  A.~Connes, M.~R.~Douglas and A.~Schwarz,
  ``Noncommutative geometry and matrix theory: Compactification on tori,''
  JHEP {\bf 9802} (1998) 003
  , hep-th/9711162.
  %%CITATION = HEP-TH 9711162;%%


\bibitem{DoHu}
  M.~R.~Douglas and C.~M.~Hull,
  ``D-branes and the noncommutative torus,''
  JHEP {\bf 9802} (1998) 008
  , hep-th/9711165.
  %%CITATION = HEP-TH 9711165;%%


\bibitem{SeWi}
N.~Seiberg and E.~Witten,
``String theory and noncommutative geometry,''
JHEP {\bf 9909} (1999) 032,
hep-th/9908142.
%%CITATION = HEP-TH 9908142;%%







\bibitem{ChWe}
  X.~Chen and E.~J.~Weinberg,
  ``ADHMN boundary conditions from removing monopoles,''
  Phys.\ Rev.\ D {\bf 67} (2003) 065020,
  hep-th/0212328.
  %%CITATION = HEP-TH 0212328;%%









%\bibitem{Myers}
%R.C. Myers,
%``Dielectric-Branes,''
%JHEP {\bf 9912} (1999) 022,
%hep-th/9910053.


%\bibitem{Connes}
%A. Connes,
%``Noncommutative geometry,'' Academic Press, 1994.\\
%See also,\\
% A. Connes, ``A Short Survey of Noncommutative Geometry,''
%hep-th/0003006.\\
%A. Connes, ``Noncommutative Geometry Year 2000,''
%math.QA/0011193.






%\bibitem{TaTe}
%  T.~Takayanagi and S.~Terashima,
%  ``c = 1 matrix model from string field theory,''
%  hep-th/0503184.
%  %%CITATION = HEP-TH 0503184;%%






%\bibitem{deWit}
%  B.~de Wit, J.~Hoppe and H.~Nicolai,
%  ``On The Quantum Mechanics Of Supermembranes,''
%  Nucl.\ Phys.\ B {\bf 305} (1988) 545.
%  %%CITATION = NUPHA,B305,545;%%









%\bibitem{KMM1}
%D.~Kutasov, M.~Marino and G.~W.~Moore, ``Some exact results on
%tachyon condensation in string field theory,'' JHEP {\bf 0010}
%(2000) 045 , hep-th/0009148.
%%%CITATION = HEP-TH 0009148;%%


%\bibitem{GeSh}
%  A.~A.~Gerasimov and S.~L.~Shatashvili,
%  ``On exact tachyon potential in open string field theory,''
%  JHEP {\bf 0010} (2000) 034
%  , hep-th/0009103.
%  %%CITATION = HEP-TH 0009103;%%









%\bibitem{Ha}
%  K.~Hashimoto,
%  ``The shape of non-Abelian D-branes,''
%  JHEP {\bf 0404} (2004) 004
%  , hep-th/0401043.




%\bibitem{HaTe}
%K.~Hashimoto and S.~Terashima, ``Boundary string field theory as a
%field theory: Mass spectrum and interaction,'' JHEP {\bf 0410}
%(2004) 040, hep-th/0408094.
%%%CITATION = HEP-TH 0408094;%%


%\bibitem{TaRa}
%  W.~I.~Taylor and M.~Van Raamsdonk,
%  ``Multiple D0-branes in weakly curved backgrounds,''
%  Nucl.\ Phys.\ B {\bf 558} (1999) 63
%  , hep-th/9904095;
%%\bibitem{Taylor:1999pr}
%%  W.~I.~Taylor and M.~Van Raamsdonk,
%  ``Multiple Dp-branes in weak background fields,''
%  Nucl.\ Phys.\ B {\bf 573} (2000) 703
%  , hep-th/9910052.








%\bibitem{spectral}
%  A.~H.~Chamseddine and A.~Connes,
%  ``The spectral action principle,''
%  Commun.\ Math.\ Phys.\  {\bf 186} (1997) 731
%  , hep-th/9606001.


%\bibitem{sigma}
%  E.~S.~Fradkin and A.~A.~Tseytlin,
%  ``Nonlinear Electrodynamics From Quantized Strings,''
%  Phys.\ Lett.\ B {\bf 163} (1985) 123;
%  %%CITATION = PHLTA,B163,123;%%
%O.~D.~Andreev and A.~A.~Tseytlin,
% ``Partition Function Representation For The Open Superstring
% Effective Action:
%Cancellation Of Mobius Infinities And
%Derivative Corrections To Born-Infeld
%Lagrangian,'' Nucl.\ Phys.\ B {\bf 311} (1988) 205.
%%%CITATION = NUPHA,B311,205;%%



%\bibitem{Is2}
%N.~Ishibashi,
%``A relation between commutative and noncommutative
%descriptions of  D-branes,''
%hep-th/9909176.
%%%CITATION = HEP-TH 9909176;%%

%\bibitem{Co}
%L.~Cornalba,
%``D-brane physics and noncommutative Yang-Mills theory,''
%Adv.\ Theor.\ Math.\ Phys.\  {\bf 4} (2000) 271,
%hep-th/9909081.
%%%CITATION = HEP-TH 9909081;%%


%\bibitem{Ok1}
%K. Okuyama,
%``A Path Integral Representation of the Map between
%Commutative and Noncommutative Gauge Fields,''
%JHEP {\bf 0003} (2000) 016,
%hep-th/9910138.
%%%CITATION = HEP-TH 9910138;%%


%\bibitem{Kara}
%%\bibitem{Castelino:1997rv}
%  J.~Castelino, S.~M.~Lee and W.~I.~Taylor,
%  %``Longitudinal 5-branes as 4-spheres in matrix theory,''
%  Nucl.\ Phys.\ B {\bf 526} (1998) 334
%  , hep-th/9712105;
%  %%CITATION = HEP-TH 9712105;%%
%  D.~Karabali, V.~P.~Nair and S.~Randjbar-Daemi,
%  ``Fuzzy spaces, the M(atrix) model and the quantum Hall effect,''
%  hep-th/0407007.
%  %%CITATION = HEP-TH 0407007;%%


%\bibitem{Hy1}
%  Y.~Hyakutake,
%  ``Torus-like dielectric D2-brane,''
%  JHEP {\bf 0105} (2001) 013
%  , hep-th/0103146.
%  %%CITATION = HEP-TH 0103146;%%

%\bibitem{bdry}
%For a review,
%P. Di Vecchia and A. Liccardo,
%``D-branes in string theory, I,''
%hep-th/9912161.
%%%CITATION = HEP-TH 9912161;%%
%
%
%\bibitem{Callan et al}
%C.~G.~Callan, C.~Lovelace, C.~R.~Nappi and S.~A.~Yost,
%``Adding Holes And Crosscaps To The Superstring,''
%Nucl.\ Phys.\ B {\bf 293} (1987) 83;
%%%CITATION = NUPHA,B293,83;%%
%``Loop Corrections to Superstring Equations of Motion,''
%Nucl.\ Phys.\ B {\bf 308} (1988) 221.
%%%CITATION = NUPHA,B308,221;%%








%\bibitem{BSFT}
%E.~Witten, ``On background independent open string field theory,''
%Phys.\ Rev.\ D {\bf 46} (1992) 5467 , hep-th/9208027;
%%%CITATION = HEP-TH 9208027;%%
%``Some Computations in Background Independent Open-String Field
%Theory,'' Phys.\ Rev.\ D {\bf 47} (1993) 3405
%, hep-th/9210065;
%%%CITATION = HEP-TH 9210065;%%

%S.~L.~Shatashvili, ``Comment on the background independent open
%string theory,'' Phys.\ Lett.\ B {\bf 311} (1993) 83;
%, hep-th/9303143;
%%%CITATION = HEP-TH 9303143;%%
%``On the problems with background independence in string theory,''
%Alg.\ Anal.\  {\bf 6}, 215 (1994) , hep-th/9311177.
%%%CITATION = HEP-TH 9311177;%%









%\bibitem{AsKi}
%  T.~Asakawa and I.~Kishimoto,
%  ``Comments on gauge equivalence in noncommutative geometry,''
%  JHEP {\bf 9911} (1999) 024
%  , hep-th/9909139.


%\bibitem{OkTe}
%  Y.~Okawa and S.~Terashima,
%  ``Constraints on effective Lagrangian of D-branes from non-commutative  gauge
%  theory,''
%  Nucl.\ Phys.\ B {\bf 584} (2000) 329
%  , hep-th/0002194.
%  %%CITATION = HEP-TH 0002194;%%

%\bibitem{Ok}
%  Y.~Okawa,
%  ``Derivative corrections to Dirac-Born-Infeld Lagrangian and  non-commutative
%  gauge theory,''
%  Nucl.\ Phys.\ B {\bf 566} (2000) 348
%  , hep-th/9909132;
%  %%CITATION = HEP-TH 9909132;%%
%%\bibitem{Te2}
%  S.~Terashima,
%  ``On the equivalence between noncommutative and ordinary gauge theories,''
%  JHEP {\bf 0002} (2000) 029
%  , hep-th/0001111;
%  %%CITATION = HEP-TH 0001111;%%
%%\bibitem{Terashima:2000ej}
%%  S.~Terashima,
%  ``The non-Abelian Born-Infeld action and noncommutative gauge theory,''
%  JHEP {\bf 0007} (2000) 033
%  , hep-th/0006058.
%  %%CITATION = HEP-TH 0006058;%%








%\bibitem{Ok2}
%K. Okuyama,
%``Boundary States in B-Field Background,''
%Phys. Lett. {\bf B499} (2001) 305,
%hep-th/0009215.
%%%CITATION = HEP-TH 0009215;%%



%\bibitem{LNT}
%F.~Larsen, A.~Naqvi and S.~Terashima, ``Rolling tachyons and
%decaying branes,'' JHEP {\bf 0302} (2003) 039
%, hep-th/0212248.
%%%CITATION = HEP-TH 0212248;%%


%\bibitem{Bigatti:1999iz}
%  D.~Bigatti and L.~Susskind,
%  %``Magnetic fields, branes and noncommutative geometry,''
%  Phys.\ Rev.\ D {\bf 62} (2000) 066004
%  , hep-th/9908056.
  %%CITATION = HEP-TH 9908056;%%


%\bibitem{Hyakutake:2003px}
%  Y.~Hyakutake,
%  %``Notes on the construction of the D2-brane from multiple D0-branes,''
%  Nucl.\ Phys.\ B {\bf 675} (2003) 241
%  , hep-th/0302190.
%  %%CITATION = HEP-TH 0302190;%%






%\bibitem{IKKT}
%N. Ishibashi, H. Kawai, Y. Kitazawa and A. Tsuchiya,
%``A Large-N Reduced Model as Superstring,''
%Nucl. Phys. {\bf B498} (1997) 467,
%hep-th/9612115.

%\bibitem{DiVeVe}
%R. Dijkgraaf, E. Verlinde and H. Verlinde,
%``Matrix String Theory,''
%Nucl. Phys. {\bf B500} (1997) 43,
%hep-th/9703030.







%\bibitem{ZS}
%M.~Marino, ``On the BV formulation of boundary superstring field
%theory,'' JHEP {\bf 0106} (2001) 059 , hep-th/0103089;
%%%CITATION = HEP-TH 0103089;%%

%V.~Niarchos and N.~Prezas, ``Boundary superstring field theory,''
%Nucl.\ Phys.\ B {\bf 619} (2001) 51 , hep-th/0103102.
%%%CITATION = HEP-TH 0103102;%%



%\bibitem{Tseysigma}
%A.~A.~Tseytlin, ``Sigma model approach to string theory effective
%actions with tachyons,'' J.\ Math.\ Phys.\  {\bf 42} (2001) 2854
%, hep-th/0011033.
%%%CITATION = HEP-TH 0011033;%%








%\bibitem{MV}
%J.~McGreevy and H.~Verlinde, ``Strings from tachyons: The c = 1
%matrix reloaded,'' JHEP {\bf 0312}, 054 (2003)
%, hep-th/0304224.
%%%CITATION = HEP-TH 0304224;%%



%\bibitem{KMS}
%I.~R.~Klebanov, J.~Maldacena and N.~Seiberg, ``D-brane decay in
%two-dimensional string theory,'' JHEP {\bf 0307}, 045 (2003)
%, hep-th/0305159.
%%%CITATION = HEP-TH 0305159;%%



%\bibitem{MVT}
%J.~McGreevy, J.~Teschner and H.~Verlinde, ``Classical and quantum
%D-branes in 2D string theory,'' JHEP {\bf 0401}, 039 (2004)
%, hep-th/0305194.
%%%CITATION = HEP-TH 0305194;%%
















%\bibitem{BaLe2}
%  D.~Bak and K.~M.~Lee,
%  ``Supertubes connecting D4 branes,''
%  Phys.\ Lett.\ B {\bf 544} (2002) 329
%  , hep-th/0206185.
%  %%CITATION = HEP-TH 0206185;%%


%\bibitem{KiLe}
%  S.~Kim and K.~M.~Lee,
%  ``Dyonic instanton as supertube between D4 branes,''
%  JHEP {\bf 0309} (2003) 035
%  , hep-th/0307048.
%  %%CITATION = HEP-TH 0307048;%%

%\bibitem{ChEtHa}
%  H.~Y.~Chen, M.~Eto and K.~Hashimoto,
%  ``The shape of instantons: Cross-section of supertubes and dyonic
%  instantons,''
%  hep-th/0609142.
%  %%CITATION = HEP-TH 0609142;%%












\end{thebibliography}
\end{document}